\documentclass[aps,prc,letterpaper,11pt,twoside,tightenlines,nofootinbib,showpacs,preprint]{revtex4}
\usepackage{graphicx}
\usepackage[sort&compress]{natbib}
\usepackage{subfigure}
\usepackage{amsmath}
\usepackage{amsfonts}
\usepackage{cancel}

\newcommand{\mev}{\textrm{ MeV}}
\newcommand{\be}{\begin{equation}}
\newcommand{\ee}{\end{equation}}
\newcommand{\ba}{\begin{eqnarray}}
\newcommand{\ea}{\end{eqnarray}}

\begin{document}

\title{Three methods to detect the predicted $D \bar{D}$ scalar meson $X(3700)$}

\author{C. W. Xiao $^1$ and E. Oset $^1$}

\affiliation{$^1$Departamento de F\'{\i}sica Te\'orica and IFIC, Centro Mixto Universidad \\de Valencia-CSIC, Institutos de Investigaci\'on de Paterna, Apartado 22085, 46071 Valencia, Spain}

\date{\today}

\begin{abstract}
In analogy to the $f_0(500)$, which appears as a $\pi \pi $ resonance in chiral unitary theory, and the $f_0(980)$, which appears as a quasibound $K \bar K$ state, the extension of this approach to the charm sector also predicts a quasibound $D \bar D $ state with mass around 3720 MeV, named as $X(3700)$, for which some experimental support is seen in the $e^+ e^- \to J/\psi D \bar D $ reaction close to the $D \bar D $ threshold. In the present work we propose three different experiments to observe it as a clear peak. The first one is the radiative decay of the $\psi(3770)$, $\psi(3770) \to \gamma X(3700) \to \gamma \eta \eta '$. The second one proposes the analogous reaction 
$\psi(4040) \to \gamma X(3700) \to \gamma \eta \eta '$ and the third reaction is the $e^+ e^- \to J/\psi X(3700) \to J/\psi \eta \eta '$. Neat peaks are predicted for all the reactions and the calculated rates are found within measurable range in present facilities.
\end{abstract}


\maketitle

\section{Introduction}

The use of the chiral unitary approach in the meson-meson interaction gives rise to the $f_0(500)(or\ \sigma),\ f_0(980),\ a_0(980)$ scalar mesons \cite{npa, norbert, markushin,rios1,rios2,juanclassic,juanenrique} from the unitarization in coupled channels of the meson-meson interaction provided by the chiral Lagrangians \cite{weinberg, gasser}. The $f_0(500)$ appears basically as a $\pi \pi$ resonance and the $f_0(980),\ a_0(980)$ as basically $K \bar{K}$ quasibound states that decay into $\pi \pi$ and $\pi \eta$ respectively. The similarity between $\bar{K}$ and $D$ ($K$ and $\bar{D}$) suggest that there could be also a $D \bar{D}$ quasibound state around $3700\mev$, that we shall call $X(3700)$, decaying into pairs of light pseudoscalars, $\pi \pi,\ \eta \eta,\ \eta \eta',\ K \bar{K}$. In an extrapolation of the chiral unitary approach to the SU(4) sector \cite{danione} it was found that, indeed, a quasibound scalar $D \bar{D}$ state with $I=0$ emerged with a small width, since transition  matrix elements from $D \bar{D}$ to the light sector were strongly suppressed. This finding has been corroborated recently in \cite{manolito,carlos} using models that incorporate heavy quark symmetry.

Later on it was found in \cite{daniee} that the bump in the $D \bar{D}$ spectrum close to the $D \bar{D}$ threshold observed at Belle in the $e^+ e^- \to J/\psi D \bar{D}$ reaction \cite{eebelle} was better interpreted in terms of the bound state below threshold, with $M_X \simeq 3723 \mev$ than with a new resonance as suggested in \cite{eebelle}. So far, this is the strongest experimental support for this state, in spite of the fact that some other reaction has been suggested to observe it. Indeed, in \cite{danirad} a suggestion was made to detect the state in the radiative decay of the $\psi(3770)$. The idea is based in the fact that the $\psi(3770)$ couples strongly to $D \bar{D}$ and with the emission of a photon one can bring the $D \bar{D}$ state below the threshold into the region of the resonance. A width of $\Gamma_{\psi \to \gamma X} = (1.05 \pm 0.41) \textrm{ KeV}$ was found which would be in the measurable range. However, a problem of this suggestion is that this peak would have to be seen over a background of $\psi \to \gamma\ + $ anything, which is estimated to have a branding ratio of the order of $10^{-2}$, judging by the rate of some measured channels reported in the Particle Date Group (PDG) \cite{pdg}, while $\Gamma_{\psi \to \gamma X} / \Gamma_{\psi} \simeq 4 \times 10^{-5}$. The signal would be of the order of 1\% of smaller on top of a background and the prospects to see it there would be dim. There is another problem since the peak for the decay appears at small photon momentum where there would be radiative decays displaying Bremsstrahlung of the photons, with accumulated strength at low photon energies, precisely where the peak of the $X$ would appear. The selection of a particular decay channel where the background would be much reduced would be then much welcome and this is what we do here, suggesting the $\eta \eta'$ channel for reasons that would be clear later on.

On the other hand, with the advent of BES\uppercase\expandafter{\romannumeral3} the production of the $\psi(4040)$ state is being undertaken and in this case the photon has more energy in the radiative decay, removing the peak from the Bremsstrahlung region, with obvious advantages.

We have also investigated another method, taking the same reaction as performed in \cite{eebelle} but looking for $e^+ e^- \to J/\psi \eta \eta'$. We predict a peak for $\eta \eta'$ production and compare the strength of the peak with the cross sections measured in \cite{eebelle} for the $J/\psi D \bar{D}$ production.

With all these studies we find out three methods which would allow to see the neat peak for that state and the widths or cross sections are found within present measuring range, such that devoted experiments would be most opportune.

\section{Formalism}

In \cite{danirad} the radiative decay of $\psi(3770)$ into $\gamma X(3700)$ was studied. The work of \cite{danione} was redone including the channels $D^+ D^-,\ D^0 \bar{D}^0,\ D^+_s D^-_s,\ \pi^+ \pi^-,\ K^+ K^-,\ \pi^0 \pi^0,\ K^0 \bar{K}^0,\ \eta \eta,\ \eta \eta',\ \eta' \eta',\ \eta_c \eta,\\ \ \eta_c \eta'$. Using a potential derived from an SU(4) extension of the SU(3) chiral Lagrangians \cite{weinberg,gasser} with an explicit SU(4) breaking for terms exchanging charm, the Bethe-Salpeter equations were solved to obtain the scattering matrix
\be 
T = [1-VG]^{-1} V,
\ee
with $V$ the potential and $G$ the loop function for the integral of intermediate two meson propagators. With this formalism a pole was obtained for the $T$ matrix around $3722\mev$ below the $D \bar{D}$ threshold. What is of relevance for the present work is the coupling of this state to the different channels. In Table \ref{tabcoupling} we show the results obtained in \cite{danirad}:
\begin{table}[htb]
\centering
\caption{Coupling of the pole at $(3722-i18)\mev$ to the channels.}\label{tabcoupling}
\begin{tabular}{|c|c|c|c|}
\hline
channel  &  Re($g_X$) [MeV]  &  Im($g_X$) [MeV]  &  $|g_X|$ [MeV]  \\
\hline
$\pi^+ \pi^-$  &  9  &  83  &  84  \\
\hline
$K^+ K^-$  &  5  &  22  &  22  \\
\hline
$D^+ D^-$  &  5962  &  1695  &  6198  \\
\hline
$\pi^0 \pi^0$  &  6  &  83  &  84  \\
\hline
$K^0 \bar{K}^0$  &  5  &  22  &  22  \\
\hline
$\eta \eta$  &  1023  &  242  &  1051  \\
\hline
$\eta \eta'$  &  1680  &  368  &  1720  \\
\hline
$\eta' \eta'$  &  922  &  -417  &  1012  \\
\hline
$D^0 \bar{D}^0$  &  5962  &  1695  &  6198  \\
\hline
$D^+_s D^-_s$  &  5901  &  -869  &  5965  \\
\hline
$\eta_c \eta$  &  518  &  659  &  838  \\
\hline
$\eta_c \eta'$  &  405  &  9  &  405  \\
\hline
\end{tabular}
\end{table}
As we can see, the largest couplings are for $D \bar{D}$ and $D_s \bar{D}_s$. However, the separation in energy of the $D_s \bar{D}_s$ component makes the $D \bar{D}$ component to stand as the more relevant meson-meson component of this state, which qualifies approximately as a $D \bar{D}$ quasibound state. The width obtained from the decay of this state in all the allowed channels is $36\mev$. Note that the transition to light, open, channels is suppressed and this determines the large lifetime of the state. We observe from the Table that the largest coupling to the light channels is to $\eta \eta'$ which also contains two different particles, hence, this will be the channel that we will adopt to have the $X(3700)$ state detected.

\begin{figure}
\centering
\includegraphics[scale=0.3]{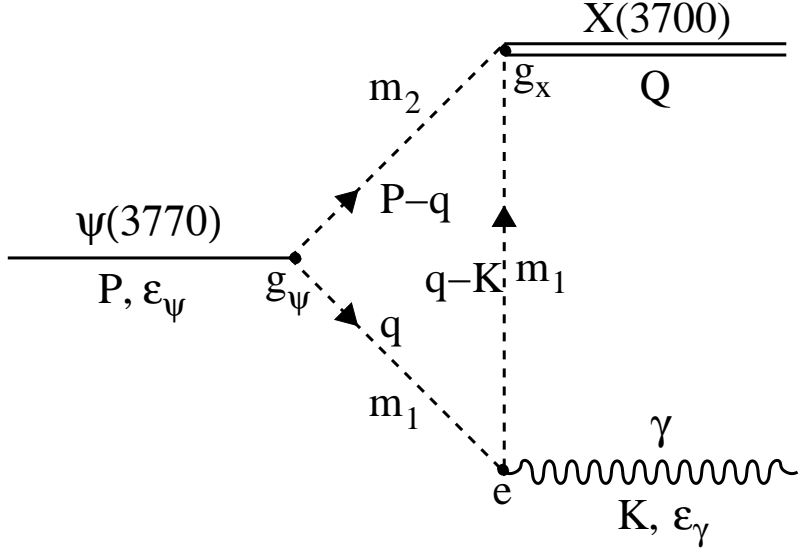}
\caption{Diagram for $\psi(3770) \to \gamma X$ that contains the $d$ term.} \label{fig1}
\end{figure}

In \cite{danirad} the decay of $\psi(3770)$ to $\gamma X$ was evaluated recalling that the $\psi(3770)$ decays basically with $D \bar{D}$. This allows one to obtain the coupling of $\psi(3770)$ to $D^+ D^-$ and then, from the triangular diagram of Fig. \ref{fig1}, the $\psi(3770) \to \gamma X$ transition
amplitude was evaluated. We do not repeat here the steps of the calculations in \cite{danirad} and quote the final results. The transition amplitude for the diagram of Fig. \ref{fig1} is given by
\be
i{\cal M} = i\epsilon_\psi^\mu(P) \epsilon_\gamma^\nu(K) {\cal T}_{\mu\nu},
\ee
and since the problem has two independent four momenta, by Lorentz invariance one may write
\be
{\cal T}_{\mu\nu} = ag_{\mu\nu}+bP_\mu P_\nu+cP_\mu K_\nu+dP_\nu K_\mu+eK_\mu K_\nu.
\ee
Due to gauge invariance only the structures $a g_{\mu\nu}$ and $d P_\nu K_\mu$ (with $P,\ K$ the $\psi$ and $\gamma$ momentum), which leads to a convergent integral, survive, and, in addition, one has $a = -d\,K\cdot P$. The $d$ coefficient is evaluated using the Feynman parameterization of the loop function corresponding to the diagram of Fig. \ref{fig1} and one finds
\be
d=-\sum_j \frac{g_\psi g_{X,j} e}{2\pi^2} \int_0^1 dx \int_0^x dy \frac{y(1-x)}{s+i\epsilon}, \label{eqd}
\ee
with $s$ given by
\be
s = (1-x)(x M_\psi^2 - m_2^2 - 2yP\cdot K)-x m_1^2,
\ee
with $e$ the electron charge ($e^2/4\pi=\alpha=1/137$), $g_\psi$ the coupling of $\psi(3770)$ to $D^+ D^-,\ g_\psi = 11.7$, and $j$ summing over the two relevant channels $D^+ D^-$ and $D^+_s D^-_s$. The partial decay width for $\psi(3770) \to \gamma X$ is given by
\be
\Gamma_{\psi\rightarrow \gamma X} = \frac{|\vec{K}|}{12\pi M_\psi^2} \; (P \cdot K)^2 \;|d|^2.
\ee
The result obtained in \cite{danirad} is \footnote{Within uncertainties it turned out to be $\Gamma_{\psi\rightarrow \gamma X} = (1.05 \pm 0.41) \textrm{ KeV}$. We refer to the value $0.65 \textrm{ KeV}$ here obtained with the standard parameters for comparison reasons.}
\be 
\Gamma_{\psi\rightarrow \gamma X} = 0.65 \textrm{ KeV}. \label{gama4a}
\ee

\begin{figure}
\centering
\includegraphics[scale=0.8]{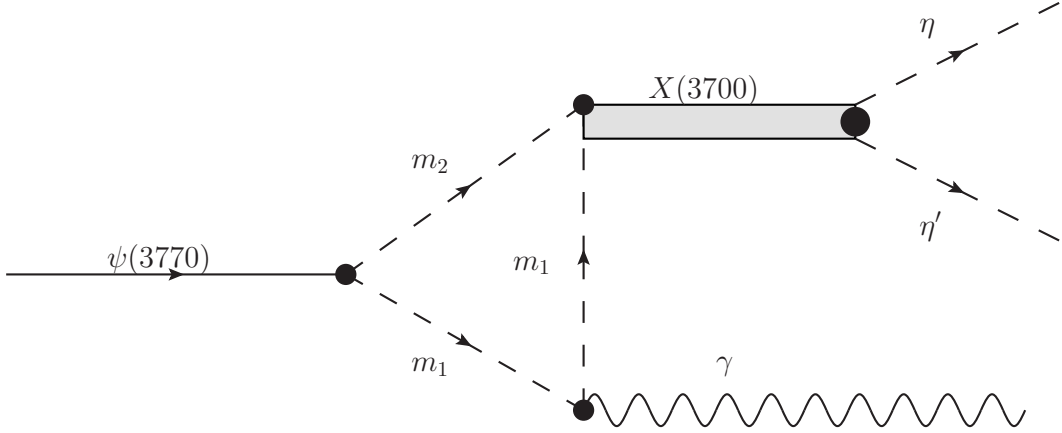}
\caption{Diagram for $\psi(3770) \to \gamma X \to \gamma \eta \eta'$.} \label{fig2}
\end{figure}

As mentioned in the Introduction, determining the peak corresponding to this process over a background of $\gamma X$ events is problematic and thus we choose the $\eta \eta'$ to detect the $X$ peak. For this the diagram of Fig. \ref{fig1} has to be changed to the one of Fig. \ref{fig2}.
Technically all we have to do is substitute $d$ by $d\,'$ where
\be 
d\,' = d \frac{1}{M_{inv}^2 - M_X^2 + i M_X \Gamma_X} g_{X,\eta \eta'},
\ee
with $d$ defined in Eq. \eqref{eqd},
\be 
M_{inv}^2 = (p_\eta + p_{\eta'})^2,
\ee
and $g_{X,\eta \eta'}$ the coupling of the $X$ to the $\eta \eta'$ channel given in Table \ref{tabcoupling}.

The relevant magnitude now is
\be 
\frac{d\Gamma}{dM_{inv}} = \frac{1}{4 (2 \pi)^3} \frac{1}{M_\psi^2} \, p_\gamma \, \tilde{p}_\eta \, \overline{\sum} \sum |T|^2,
\ee
which provides the invariant mass distribution, where
\ba 
p_\gamma &=&  \frac{\lambda^{1/2}(M_\psi^2, 0, M_{inv}^2)}{2 M_\psi}, \\
\tilde{p}_\eta &=& \frac{\lambda^{1/2}(M_{inv}^2, m_\eta^2, m_{\eta'}^2)}{2M_{inv}}, \\
\overline{\sum} \sum |T|^2 &=& \frac{2}{3} \, |d\,'|^2 (K\cdot P)^2,
\ea
with $p_\gamma$, $\tilde{p}_\eta$ the $\gamma$ momentum in the $\psi(3770)$ rest frame and the $\eta$ momentum in the $\eta \eta'$ rest frame respectively.

\begin{figure}
\centering
\includegraphics[scale=0.8]{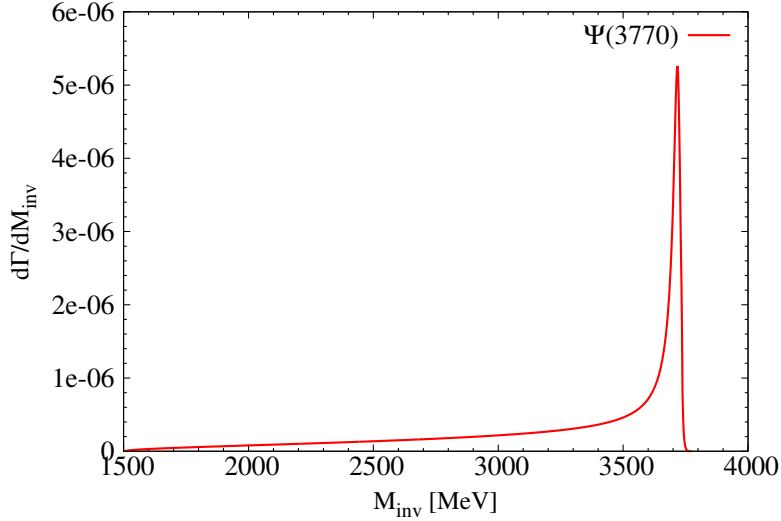}
\caption{The mass distribution of the $\eta \eta'$ in the decay of $\psi(3770)$ to $\gamma X(3700) \to \gamma \eta \eta'$.} \label{fig3}
\end{figure}

In Fig. \ref{fig3} we show this distribution,
and we see a clear peak at $M_{inv} \simeq 3722 \mev$ with a narrow width. The peak is still around the upper threshold for the invariant mass. However, the fact that we have chosen a neutral channel to identify the $X$ state prevents Bremsstrahlung to occur and the identification of a peak there would be a clear signal of a state. The integrated width around the peak ($3600 < M_{inv} < 3770 \mev$) gives
\be 
\Gamma = \int_{3600}^{3770} \frac{d\Gamma}{dM_{inv}} dM_{inv} = 0.293 \textrm{ KeV}, \label{eqga}
\ee
which is smaller than the total width of Eq. \eqref{gama4a} which integrates over the whole range of $M_{inv}$.

The largest contribution comes from the $D^+ D^-$ channel which by itself provides 73\% of the rate. The coherent sum with the $D_s^+ D_s^-$ contribution makes up for the rest of the rate.

The width of Eq. \eqref{eqga} represents a branching ratio of $10^{-5}$. In this sense one should note that CLEO has set thresholds of the order of magnitude of $10^{-4}$ and at BES\uppercase\expandafter{\romannumeral3} one can get a production $\psi(3770)$ of about a factor one hundred times bigger, which would make this measurement feasible in that Lab.

\section{Radiative decay of the $\psi(4040)$}

\begin{figure}
\centering
\includegraphics[scale=0.8]{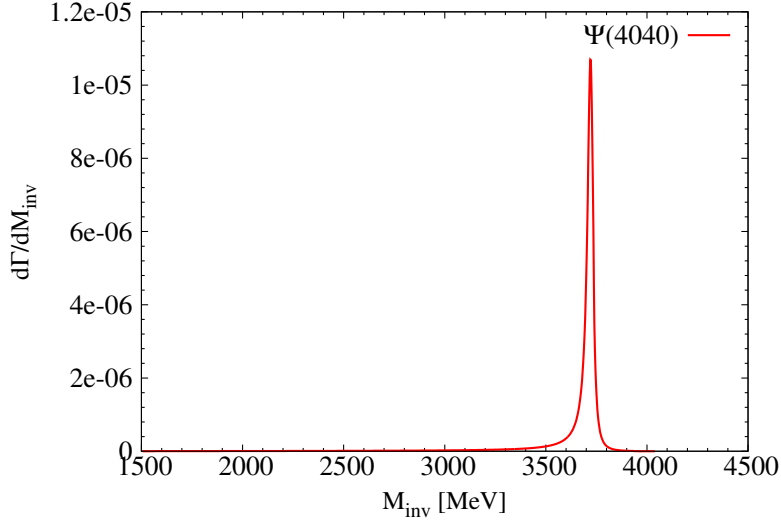}
\caption{The mass distribution of $\eta \eta'$ in the $\psi(4040)$ decay to $\gamma X(3700) \to \gamma \eta \eta'$.} \label{fig4}
\end{figure}

The $\psi(4040)$ shares the same quantum numbers as the $\psi(3770)$, however the largest branching ratio is not to $D \bar{D}$ but to $D^* \bar{D} + cc$. From the data in the PDG we find that
\ba 
\frac{\Gamma(D \bar{D})}{\Gamma(D^* \bar{D} + cc)} &=& 0.24 \pm 0.05 \pm 0.12, \\
\frac{\Gamma(D^* \bar{D}^*)}{\Gamma(D^* \bar{D} + cc)} &=& 0.18 \pm 0.14 \pm 0.03,
\ea
Assuming that the $D^* \bar{D} + cc$, $D \bar{D}$ and $D^* \bar{D}^*$ provide most of the contribution, this allows us to get the coupling
\be 
g_{\psi(4040), D^+ D^-} = 2.15, \label{eqcoup}
\ee
and then we can recalculate the invariant mass distribution and width for $\psi(4040) \to \gamma X(3700)$. In Fig. \ref{fig4} we show the results for the invariant mass distribution.
We find now a neat peak around the mass of the $X$. The novelty here is that the peak is far away from all thresholds which could eventually be seen in the spectrum of inclusive $d\Gamma / dE_\gamma$ without the risk to confuse the peak with Bremsstrahlung like in the case of the $\psi(3770) \to \gamma X$. In any case, as advocated here, the direct measurement of the $\eta \eta'$ channel should drastically reduce the background and allow a clear peak to be identified. The integrated width around this peak ($3600 < M_{inv} < 3800 \mev$) gives
\be 
\Gamma = \int_{3600}^{3800} \frac{d\Gamma}{dM_{inv}} dM_{inv} = 0.496 \textrm{ KeV}, \label{eqga2}
\ee
which is about double than in the case of the $\psi(3770)$. In this case, the larger phase space for decay has overcome the reduction due to the reduced coupling of Eq. \eqref{eqcoup}.

We should note that the largest contribution comes from the $D^+ D^-$ channel, this channel alone providing about half the rate of Eq. \eqref{eqga2} while $D_s^+ D_s^-$ alone only given 19\% of this rate.

The width of Eq. \eqref{eqga2} is a bit bigger than the one obtained for the $\psi(3770)$, yet, the rate of production at BESIII is smaller. Present plans are to produce 2.8 million $\psi(4040)$ events and no plans are made for the future yet \footnote{We would like to thank Cheng-Ping Shen for providing us the information.}. With this statistics and the width of Eq. \eqref{eqga2}, which corresponds to a branching ratio of $6.2 \times 10^{-6}$, one could get about 17 events of this radiative decay. It is clear that more statistics would be needed to see a clear peak.

\section{The $e^+ e^- \to J/\psi X \to J/\psi \eta \eta'$ reaction}

\begin{figure}
\centering
\includegraphics[scale=0.3]{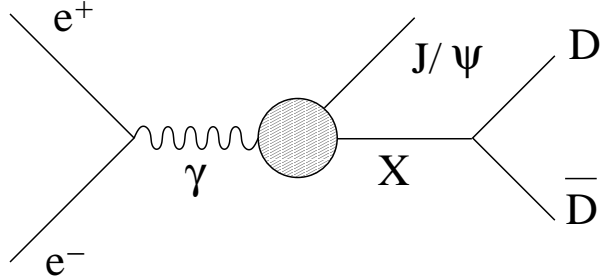}
\caption{Feynman diagram of the reaction $e^+ e^- \to J/\psi X \to J/\psi D \bar{D}$.} \label{fig5}
\end{figure}

In \cite{daniee} the $e^+ e^- \to J/\psi X \to J/\psi D \bar{D}$ reaction was studied and it was concluded that the data on the $D \bar{D}$ invariant mass distribution was better described in terms of the $X(3700)$ resonance that in terms of a new state suggested in \cite{eebelle}. The mechanism for this reaction is given in Fig. \ref{fig5}. 
The differental cross section is given by \cite{daniee}
\be 
\frac{d\sigma}{dM_{inv}(D \bar{D})} = \frac{1}{(2\pi)^3} \frac{m_e^2}{s \sqrt{s}} |\vec{k}|\; |\vec{p}|\; |T|^2,
\ee
with
\ba 
|\vec{k}| &=& \frac{\lambda^{1/2}(M_{inv}^2(D \bar{D}), m_D^2, m_D^2)}{2M_{inv}(D \bar{D})}, \\
|\vec{p}| &=& \frac{\lambda^{1/2}(s, M_{J/\psi}^2, M_{inv}^2(D \bar{D}))}{2\sqrt{s}},
\ea
where $T$ is given by
\be 
T = C \frac{1}{M_{inv}^2(D \bar{D}) - M_X^2 + i \Gamma_X M_X}. \label{eqT}
\ee
As in \cite{daniee} we restrain from giving absolute values but we can give relative values with respect to $D \bar{D}$ production simply multiply $T$ of Eq. \eqref{eqT} by $g_{X, \eta \eta'} / \sqrt{2} g_{X, D^+ D^-}$, where the factor $\sqrt{2}$ will take into account in $|T|^2$ that we compare $\eta \eta'$ production versus $D^+ D^- + D^0 \bar{D}^0$ production.
\begin{figure}
\centering
\includegraphics[scale=0.8]{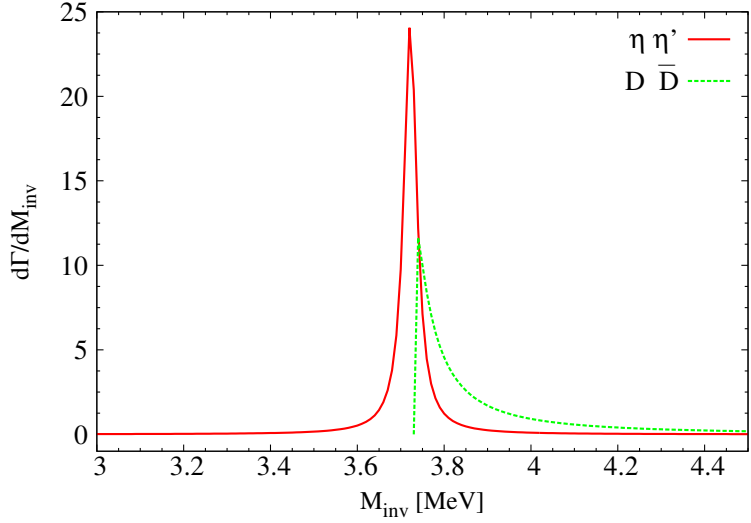}
\caption{The mass distribution of the final states $J/\psi \eta \eta'$ compared to $J/\psi D \bar{D}$.} \label{fig6}
\end{figure}
In Fig. \ref{fig6} we show the results for $\eta \eta'$ production with the same scale as for $D \bar{D}$ production. We can see that the strength of the peak is bigger for $\eta \eta'$ production than for $D \bar{D}$, in spite of having a smaller coupling to $X(3700)$. The reason is that the $\eta \eta'$ production is not suppressed by the threshold factors that inhibit $D \bar{D}$ production. The peak seen in the $\eta \eta'$ mass spectrum is neat and the strength larger than for $D \bar{D}$ production. Since $D \bar{D}$ has been observed in \cite{eebelle}, this guarantees that the $\eta \eta'$ peak is within present measurable range.

\section{Conclusions}

In the present work, we have investigated some reactions by means of which one could observe the predicted scalar meson formed as a quasibound state of $D \bar D$. This state appears in analogy to the $f_0(500)$ and $f_0(980)$ states which are described within the chiral unitary approach as a $\pi \pi $ resonance and a quasibound $K \bar K$ state respectively. Some suggestion had been made before to observe this state in the $\psi(3770) \to \gamma X(3700)$ decay by looking at the $\gamma$ energy distribution. Yet, this has the inconvenience of having to observe a small peak in a large background. In order to suppress the background we have chosen one of the main decay channels of the  $X(3700)$ state, the $\eta \eta '$ channel, and suggest to look at the $\eta \eta '$ invariant mass distribution in the reaction  $\psi(3770) \to \gamma X(3700) \to \gamma \eta \eta '$. Since BESIII already can produce the $\psi(4040)$, we also suggest to look at the $\psi(4040) \to \gamma X(3700) \to \gamma \eta \eta '$ decay channel. A third reaction was motivated by the only indirect experimental \textquoteleft \textquoteleft evidence" of this state. Indeed, in the BELLE reaction 
$e^+ e^- \to J/\psi D \bar D $ \cite{eebelle}, a peak was observed in the $D \bar D $ invariant mass distribution close to the $D \bar D $ threshold, which was interpreted in 
\cite{daniee} as a signal of a $D \bar D$ resonance below the $D \bar D $ threshold. In the present work we have suggested to look at the reaction 
$e^+ e^- \to J/\psi \eta \eta ' $, allowing the $X(3700)$ to be produced and decay into $\eta \eta ' $. 

   We find clear peaks in all the invariant mass distributions of $\eta \eta '$.
In the two radiative decays, the rates are within present measurable range at BESIII, although in the case of $\psi(4040)$ radiative decay the statistics with presently planned $\psi(4040)$ production would be very low. In the case of the $e^+ e^-$ reaction we do not evaluate absolute cross sections and we find more instructive to compare the cross section of the  $e^+ e^- \to J/\psi \eta \eta ' $ reaction with the one of $e^+ e^- \to J/\psi D \bar D $ already measured. We observe that the cross section for the former reaction is bigger than for the latter one and produces a clear peak that does not have the ambiguity of a threshold enhancement as in the $e^+ e^- \to J/\psi D \bar D$ reaction. This is the best guarantee that the reaction is within measurable range. 

    The experimental search for this state is timely and its observation would clarify issues concerning the interaction of hadrons in the charm sector, which is not so well known as the non charmed one, and which would be much welcome.

\section*{Acknowledgements}

This work is partly supported by DGICYT contract number FIS2011-28853-C02-01, and the Generalitat Valenciana in the program Prometeo, 2009/090. We acknowledge the support of the European Community-Research Infrastructure Integrating Activity Study of Strongly Interacting Matter (acronym HadronPhysics3, Grant Agreement n. 283286) under the Seventh Framework Programme of the EU.

\end{document}